\documentclass[11pt]{article}

\usepackage[margin=1in]{geometry}
\usepackage{setspace}
\usepackage{amsmath,amssymb,amsfonts}
\usepackage{bm}

\usepackage{graphicx}
\usepackage{subcaption}
\usepackage{booktabs}
\usepackage[font=small,labelfont=bf]{caption}
\graphicspath{{figures/}{images/}}

\usepackage[numbers,sort&compress]{natbib}

\usepackage[colorlinks=true,allcolors=blue,breaklinks=true]{hyperref}

\begin{document}

\title{\bfseries An SPH--mesh Coupling for Vesicle Dynamics in Shear Flow}

\author{%
Kuiliang Wang\,$^{1}$\quad
Xinwei Cai\,$^{1}$\quad
Ting Ye\,$^{2}$\quad
Xuejin Li\,$^{1}$\quad
Xin Bian\,$^{1,}$\thanks{Corresponding author:
\href{mailto:bianx@zju.edu.cn}{bianx@zju.edu.cn}}
\\[1.2ex]
\normalsize $^{1}$State Key Laboratory of Fluid Power and Mechatronic Systems,\\
\normalsize Department of Engineering Mechanics, Zhejiang University, Hangzhou 310027, China\\[0.4ex]
\normalsize $^{2}$School of Mathematics, Jilin University, Changchun 130012, China
}

\date{}

\maketitle

\begin{abstract}
\noindent
We present a novel computational framework that couples smoothed particle hydrodynamics~(SPH) with a triangulated membrane mesh to simulate the dynamics of vesicles suspended in fluids.
A novel interface-tracking approach enforces membrane impermeability naturally, without resorting to non-physical constraints such as particle reflection or bounce-back
boundary conditions. The membrane model incorporates four distinct 
bending energy formulations, namely the minimal model,
the spontaneous curvature (SC) model, the bilayer couple (BC) model, and
the area difference elasticity (ADE) model, providing a versatile tool for diverse biophysical scenarios. The framework is rigorously validated against equilibrium shapes and
tank-treading motion of a vesicle, demonstrating excellent agreement
with previous theoretical and numerical studies. A systematic investigation into the effects of each bending model on the vesicle's inclination angle, revolution frequency, and morphology in shear flow reveals key physical insights. Notably, spontaneous curvature has a negligible effect on steady-state orientation but profoundly alters rotational dynamics at low reduced volumes through the emergence of dumbbell-like shapes with deep constrictions. In contrast, the BC and ADE models induce characteristic asymmetric and stomatocyte morphologies. Our results establish the proposed SPH--mesh coupling as an accurate and robust tool for exploring the complex, shape-dependent dynamics of vesicles in fluid flows.
\end{abstract}


\vspace{1.5em}


\section{\label{sec:introduction}Introduction}

Vesicles are closed membranes composed of a phospholipid bilayer that
enclose an aqueous fluid, and they serve as a fundamental structural unit
of biological cells. As simplified models of red blood cells and other
biological entities, vesicles have attracted considerable attention
because they reproduce many of the rich shapes and dynamical behaviors
observed in living cells while remaining amenable to controlled
experiments and quantitative modeling~\cite{seifert1997configurations,
sackmann1996supported}. Understanding how vesicles deform, orient, and
move in fluid flows is therefore of great importance for applications
ranging from microfluidic manipulation and drug delivery to the rheology
of blood and other biological suspensions~\cite{danker2009vesicles,
abreu2014fluid}.

The equilibrium shapes of vesicles are governed by the bending elasticity
of the membrane. The pioneering works of Canham~\cite{canham1970minimum}
and Helfrich~\cite{helfrich1973elastic} established the curvature-energy
framework that explains the biconcave shape of red blood cells and a
variety of other morphologies. Subsequent extensions, including the
bilayer-couple (BC) model~\cite{svetina1989membrane, bozic1992role} and
the area-difference elasticity (ADE) model~\cite{ seifert1997configurations, abreu2014fluid, 
miao1994budding}, incorporate the non-local
coupling associated with the area difference between the two monolayers
of the bilayer, and successfully predict prolate, oblate, stomatocyte,
and budding shapes. Comprehensive phase diagrams of vesicle shapes based
on these models have been established both theoretically and
numerically~\cite{seifert1991shape, ziherl2005nonaxisymmetric,
bian2020bending}.

Beyond static shapes, the dynamics of vesicles in flow has been a central
topic in soft matter physics and fluid mechanics. In shear flow, vesicles
exhibit a sequence of distinct dynamical regimes, including tank-treading,
tumbling, and trembling (or vacillating-breathing) motions, depending on
the reduced volume, the viscosity ratio, and the flow
strength~\cite{kraus1996fluid, kantsler2005orientation,
kantsler2006transition, mader2006dynamics, deschamps2009dynamics}.
Kraus \textit{et al.}~\cite{kraus1996fluid} provided a
theoretical and numerical description of vesicle tank-treading in shear
flow, while later analytical theories~\cite{misbah2006vacillating,
lebedev2007dynamics, vlahovska2007dynamics, danker2007dynamics} clarified
the transitions between these regimes. These predictions have been
confirmed by elegant experiments~\cite{kantsler2005orientation,
kantsler2006transition, deschamps2009dynamics}, establishing vesicle
dynamics as a well-characterized benchmark problem for numerical methods.
The interplay between membrane bending, confinement, 
and flow gives rise to a wealth of shape transitions 
and dynamical states, as demonstrated for both fluid 
vesicles and red blood cells in capillary and shear 
flows~\cite{noguchi2005shape, fedosov2014deformation}.

A wide range of numerical methods has been developed to simulate vesicle
and capsule dynamics in fluids. Boundary-integral
methods~\cite{kraus1996fluid, pozrikidis1990axisymmetric, 
zhao2011dynamics, veerapaneni2011fast} are highly accurate for
Stokes-flow problems and have been used extensively to study
tank-treading and tumbling. The immersed boundary
method~\cite{zhu2002simulation, kim2010simulating} couples an immersed
membrane with a background fluid grid, and has become one of the most
popular approaches for fluid--structure interaction with deformable
membranes. Phase-field~\cite{biben2003tumbling, du2004phase,
biben2005phase}, level-set~\cite{salac2011level, doyeux2013simulation},
and lattice-Boltzmann~\cite{kaoui2011two, guckenberger2016bending}
methods have likewise been applied to vesicle and red-blood-cell
simulations. Many of these methods, and the corresponding numerical
challenges of imposing membrane constraints and fluid--membrane coupling,
have been actively discussed in the computational fluid dynamics
community~\cite{zhao2011dynamics, salac2011level, doyeux2013simulation,
guckenberger2016bending}. A recurring difficulty in such grid-based
approaches is the need to enforce membrane inextensibility and enclosed
volume conservation, which are often achieved through Lagrange multipliers
or penalty forces that introduce additional non-physical parameters.
Coarse-grained particle methods such as dissipative 
particle dynamics (DPD)~\cite{Li2017} and smoothed dissipative 
particle dynamics (SDPD)~\cite{Ellero2018} have proven particularly 
powerful in this context, providing quantitative 
descriptions of the mechanics, rheology, and flow 
dynamics of red blood cells and vesicles across 
scales~\cite{pivkin2008accurate, 
fedosov2010multiscale, li2012continuum, 
ye2016particle, Han2024, Amoudruz2024}. These approaches have elucidated 
the deformation and dynamics of cells in 
microchannels and complex flows~\cite{lei2013blood, 
fedosov2014multiscale, li2017biomechanics, 
li2018mechanics}, and continuum vesicle/RBC models 
have likewise revealed rich shape transitions and 
tank-treading behaviour in capillary and shear 
flows~\cite{noguchi2005dynamics}. 
Related coupling strategies have been explored for red 
blood cells, such as the hybrid smoothed dissipative 
particle dynamics–immersed boundary method of Ye \textit{et al.}, 
in which a spring-network membrane is coupled to a 
particle-based fluid to reproduce tank-treading 
and tumbling motions~\cite{ye2017hybrid}. In contrast, 
the present work couples an SPH fluid directly with 
a triangulated membrane equipped with multiple 
bending models.

Smoothed particle hydrodynamics
(SPH)~\cite{monaghan2012smoothed} is a fully Lagrangian, mesh-free method that is
particularly attractive for problems involving multiple fluid phases,
large deformations, and moving interfaces. 
Owing to its Lagrangian and mesh-free nature, 
SPH has undergone rapid methodological 
development and has been applied to a broad 
range of multiphase and fluid–structure 
interaction problems~\cite{liu2010smoothed, 
ye2019smoothed}. 
Multiphase SPH formulations~\cite{hu2006MultiphaseSPHMethod, 
morris2000simulating, zheng2019multiphase} naturally handle fluids with 
different densities and viscosities on either 
side of an interface, and the weak compressibility
of the standard SPH scheme provides an intrinsic mechanism for
approximately enforcing incompressibility. SPH has been applied to model
red blood cells and capsules by representing the membrane with bonded
particles or spring networks~\cite{hosseini2009particle,tanaka2005microscopic}. However, when the membrane itself is
represented purely by particles, accurately capturing the
curvature-dependent bending energy and preventing fluid particles from
penetrating the membrane remain challenging, and often require additional
artificial constraints.
Building on our previous SPH studies for 
the dynamics of droplets, elliptical particles, 
and squirmers in flow~\cite{wang2023dynamics, 
cai2024dynamics, cai2025simulating}, the 
present work extends the particle-based 
framework to vesicle membranes.

In this work, we propose a new method that couples a multiphase SPH
solver with a triangulated-mesh representation of the vesicle membrane. In
contrast to conventional immersed boundary methods, where the membrane
moves according to the interpolated fluid velocity, here the position of
the membrane mesh is determined by the interface between the inner and
outer fluid particles, identified through a color function. This
interface-tracking strategy ensures the impermeability of the membrane
without imposing non-physical reflection or bounce-back boundary
conditions, while the enclosed volume is preserved automatically through
the weak compressibility of the SPH method, eliminating the need for an
explicit volume constraint. Furthermore, the triangulated membrane model
incorporates four distinct bending energy formulations, namely the minimal, SC,
BC, and ADE models, allowing a unified treatment of a broad range of
vesicle morphologies and dynamical behaviors within a single framework.
We validate the method against equilibrium vesicle shapes and the
tank-treading dynamics predicted by theory, and then apply it to
investigate the inclination, revolution frequency, and special
morphologies of vesicles under shear flow for the different bending
models.

The remainder of this paper is organized as follows.
Section~\ref{sec:methods} introduces the four bending models, the
discretization of the membrane, and the coupled SPH--mesh numerical
method. Section~\ref{sec:numerical results} presents the validation of
the method and the simulation results for vesicles in shear flow under
the different bending models. Finally, Section~\ref{sec:conclusion}
summarizes the main conclusions and discusses future directions.

\section{\label{sec:methods}Theories and numerical methods}

\subsection{Theories of bending models}

The equilibrium shape and dynamical response of a vesicle are largely
dictated by the bending elasticity of its membrane. In the present study,
four bending models are considered, all of which were developed to
describe the mechanical behavior of the phospholipid
bilayer~\cite{seifert1997configurations, canham1970minimum, 
helfrich1973elastic, svetina1989membrane, bozic1992role, 
wiese1992budding}.

An early and widely used formulation was introduced by
Canham~\cite{canham1970minimum}, who expressed the bending energy of the
membrane as
\begin{equation}
E_{C}=\frac{\kappa_{b}}{2} \int\left(C_{1}^{2}+C_{2}^{2}\right) d A,
\label{Canham_energy}
\end{equation}
where $\kappa_{b}$ denotes the bending modulus, and $C_{1}$ and $C_{2}$
are the two principal curvatures of the membrane surface. Owing to its
simplicity, this formulation is commonly referred to as the
\textit{minimal model}.

Building upon this description, Helfrich~\cite{helfrich1973elastic}
accounted for the asymmetry of the bilayer in its stress-free state by
introducing a spontaneous curvature, which leads to the bending energy
\begin{equation}
E_{\mathrm{Helfrich}}=2\kappa_{b}\int(H-H_{0})^{2}dA+\kappa_{g}\int G\,dA,
\label{Helfrich_energy}
\end{equation}
where $H$ is the mean curvature, $H_{0}$ is the spontaneous curvature,
$\kappa_{g}$ is the Gaussian bending modulus, and $G$ is the Gaussian
curvature. This formulation is known as the \textit{spontaneous-curvature
(SC) model}. For the closed vesicles of spherical topology considered in
this work, the Gauss--Bonnet theorem guarantees that the last term in
Eq.~(\ref{Helfrich_energy}) remains constant and can therefore be omitted
from the energy minimization. In the special case $H_{0}=0$, the Helfrich
energy reduces to the Canham energy up to this constant Gaussian
contribution, so that the SC model recovers the minimal model.

Further refinements of the Helfrich model were subsequently proposed to
incorporate the area difference between the outer and inner monolayers of
the bilayer, giving rise to two distinct descriptions. In the first, the
monolayers are assumed to be incompressible, so that the area difference
is fixed and preserved throughout the deformation of the vesicle; this
gives the \textit{bilayer-couple (BC) model}~\cite{svetina1989membrane,
bozic1992role}. In the second, the monolayers are allowed a finite
compressibility, and the deviation of the area difference from its
preferred value is penalized through an elastic energy term; this yields
the \textit{area-difference elasticity (ADE)
model}~\cite{seifert1997configurations, wiese1992budding}. The total
bending energy of the ADE model is the sum of the Helfrich energy and the
area-difference elasticity energy,
\begin{equation}
  \begin{split}
    E &=E_{\text{Helfrich}}+E_{ADE}\\
    &=2 \kappa_{b} \int\left(H-H_{0}\right)^{2} d A
      +\frac{\alpha \kappa_{b} \pi}{2 A D^{2}}
      \left(\Delta A-\Delta A_{0}\right)^{2},
  \end{split}
\label{Total_energy}
\end{equation}
where $\alpha \kappa_{b}$ is the elastic constant associated with the
area difference, $A$ is the total membrane area, and $D$ is the thickness
of the bilayer. The area difference $\Delta A$ is related to the mean
curvature through
\begin{equation}
\Delta A=2 D \int H \, d A,
\label{Area_difference}
\end{equation}
and $\Delta A_{0}$ denotes the reference (preferred) area difference. The
BC model is recovered from Eq.~(\ref{Total_energy}) by setting $H_{0}=0$
and enforcing the hard constraint $\Delta A = \Delta A_{0}$, which
corresponds to the limit of vanishing monolayer compressibility.

\subsection{Numerical discretization of the membrane}

The four bending models introduced above can be treated within a unified
formulation. Substituting Eq.~(\ref{Area_difference}) into
Eq.~(\ref{Total_energy}) and expanding the quadratic terms, the total
energy is rewritten as
\begin{equation}
  \begin{split}
  E= &2 \kappa_{b} \int H^{2} d A-4 \kappa_{b} H_{0} \int H d A
     +2 \kappa_{b} H_{0}^{2} A\\
  &+\frac{2 \alpha \kappa_{b} \pi}{A} \left(\int H d A\right)^{2}\\
  &-\frac{2 \alpha \kappa_{b} \pi}{A} \frac{\Delta A_{0}}{D} \int H d A
    +\frac{\alpha \kappa_{b} \pi}{2 A}\left(\frac{\Delta A_{0}}{D}\right)^{2}.
  \end{split}
\label{total_energy_integral}
\end{equation}
This expression makes explicit that all four models require only two
geometric quantities to be evaluated on the discretized surface, namely
the integrated squared mean curvature $\int H^{2}\,dA$ and the integrated
mean curvature $\int H\,dA$, together with the total area $A$.

\begin{figure*}[tbp]
    \centering
    \includegraphics[width=0.95\textwidth, keepaspectratio]{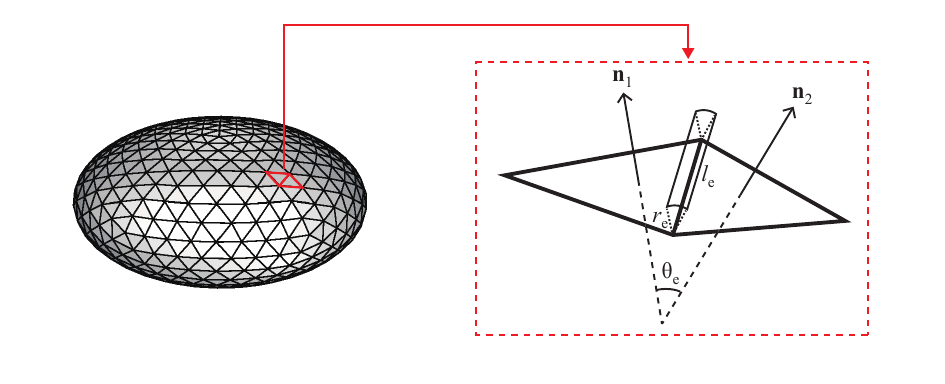}
    \caption{An oblate vesicle membrane discretized with a triangular mesh.
Each edge $e$, of length $l_{e}$, is interpreted as a portion of a
cylindrical surface of vanishing radius $r_{e}$, bent by the dihedral
angle $\theta_{e}$ between the outward normals $\mathbf{n}_{1}$ and
$\mathbf{n}_{2}$ of the two adjacent faces. The auxiliary radius $r_{e}$
cancels in the discrete energy and does not enter the final
expressions.\label{fig:mesh_edge}}
\end{figure*}

The vesicle membrane is represented by a triangulated surface, as
illustrated in Fig.~\ref{fig:mesh_edge}.
Bian \textit{et al.}~\cite{bian2020bending} systematically compared four
schemes for discretizing the bending energy on such meshes; guided by
their results, we adopt the edge-based scheme proposed by
J\"ulicher~\cite{julicher1996morphology}, which offers a good compromise
between accuracy and computational simplicity. In this scheme, each edge
of the mesh is interpreted as a portion of a cylindrical surface with a
vanishingly small radius of curvature $r_{e}\to 0$, as sketched in
Fig.~\ref{fig:mesh_edge}. On such an edge the principal curvatures are
$C_{1}=1/r_{e}$ and $C_{2}=0$, whereas the flat interior of each triangle
carries no curvature, $C_{1}=C_{2}=0$. The integrated mean curvature is
thus obtained as a sum over all $N_{e}$ edges,
\begin{equation}
 \int H \, d A=\sum_{e=1}^{N_{e}} \lim_{r_{e} \to 0}
 \frac{1}{2}\left(0+\frac{1}{r_{e}}\right) \theta_{e}\, r_{e}\, l_{e}
 =\frac{1}{2} \sum_{e=1}^{N_{e}} \theta_{e}\, l_{e},
\label{integral_sum}
\end{equation}
where $\theta_{e}$ and $l_{e}$ are the dihedral angle and the length of
edge $e$, respectively. The auxiliary radius $r_{e}$ cancels, confirming
that the limit of an infinitely sharp edge is well defined. Likewise, the
total surface area is evaluated by summing the areas of all $N_{t}$
triangular faces,
\begin{equation}
A=\sum_{t=1}^{N_{t}} A^{t}.
\label{area_sum}
\end{equation}

The same limiting procedure cannot be applied to $\int H^{2}\,dA$, because
squaring the curvature leaves an uncancelled factor $1/r_{e}$ that diverges
as $r_{e}\to 0$. We therefore transfer the curvature from the edges to the
vertices. Equation~(\ref{integral_sum}) assigns the integrated mean
curvature to individual edges; splitting each edge contribution equally
between its two end vertices yields the local mean curvature at vertex $i$,
\begin{equation}
H_{i}=\frac{1}{4 A_{i}} \sum_{e:\langle i,j\rangle} \theta_{e}\, l_{e},
\label{vertex_curvature}
\end{equation}
where the sum runs over the $N_{e}^{i}$ edges incident to vertex $i$, and
$A_{i}$ is the area attributed to that vertex, obtained by sharing each
triangle area equally among its three vertices,
\begin{equation}
A_{i}=\frac{1}{3} \sum_{t:\langle i,j,k\rangle} A^{t}.
\label{vertex_area}
\end{equation}
Since $\sum_{i} A_{i}=A$ and $\sum_{i} H_{i} A_{i}$ recovers
Eq.~(\ref{integral_sum}), this vertex-based description is consistent with
the edge-based one. The integrated squared mean curvature then follows as a
sum over all $N_{v}$ vertices,
\begin{equation}
\int H^{2} \, d A=\sum_{i=1}^{N_{v}} H_{i}^{2} A_{i}.
\label{squared_curvature_sum}
\end{equation}

Equations~(\ref{integral_sum})--(\ref{squared_curvature_sum}) express the
discrete energy~(\ref{total_energy_integral}) entirely in terms of the
dihedral angles $\theta_{e}$, the edge lengths $l_{e}$ and the triangle
areas $A^{t}$, and hence in terms of the vertex coordinates. The elastic
force acting on each vertex then follows from the negative gradient of the
energy with respect to its position,
\begin{equation}
\mathbf{F} = -\frac{\partial E}{\partial \mathbf{x}},
\label{Force_energy}
\end{equation}
which requires only the geometric derivatives
\begin{equation}
\frac{\partial \theta_{e}}{\partial \mathbf{x}}, \quad
\frac{\partial l_{e}}{\partial \mathbf{x}}, \quad
\frac{\partial A^{t}}{\partial \mathbf{x}}.
\label{partial_x}
\end{equation}
These derivatives are evaluated analytically for every edge and triangle
incident to a given vertex, so that the bending force is computed
efficiently by looping over the local mesh connectivity.

\subsection{\label{sec:SPH and IBM}Multiphase SPH method coupled with the membrane}

The fluids inside and outside the vesicle are modeled by a multiphase
smoothed particle hydrodynamics (SPH) method, while the coupling between
the membrane and the surrounding fluids is achieved through a novel
immersed-boundary strategy. The two ingredients are described in turn
below.

\subsubsection{Multiphase SPH method}

The fluid motion is governed by the isothermal Navier--Stokes equations,
written here in the Lagrangian frame,
\begin{equation}
\begin{aligned}
&\frac{d\rho }{dt} =-\rho\,\nabla \cdot \mathbf{v}, \\
&\frac{d\mathbf{v}}{dt} = \frac{1}{\rho}\left( -\nabla p
+ \mu\nabla^{2}\mathbf{v} + \mathbf{F}_{g} \right),
\end{aligned}
 \label{governing}
\end{equation}
where $\rho$, $\mathbf{v}$, $p$ and $\mu$ denote the density, velocity,
pressure and dynamic viscosity, respectively. $\mathbf{F}_{g}$ is the
body force, which is omitted in this work.

To close the system within a weakly compressible framework, the pressure
is related to the density through an artificial equation of state,
\begin{equation}
p=c_{s}^{2}\left( \rho - \rho_{\mathrm{ref}} \right),
\label{state}
\end{equation}
where $c_{s}$ is an artificial speed of sound and $\rho_{\mathrm{ref}}$ is
a reference density. Subtracting the reference density in the SPH
discretization of the pressure gradient reduces the associated numerical
error~\cite{morris2000simulating}.

In the SPH formalism the fluid is represented by a set of moving
particles that carry the flow properties. The density of a particle is
recovered by summing the mass contributions of its neighbors,
\begin{equation}
\rho_{i}=m_{i} \sum_{j}W_{ij},
\label{rhosum}
\end{equation}
where the particle mass $m_{i}$ is constant in time. The weight (kernel)
function is
\begin{equation}
W_{ij}=W\left( \mathbf{r}_{ij},h \right),
\label{kernel}
\end{equation}
with $\mathbf{r}_{ij}=\mathbf{r}_{i}-\mathbf{r}_{j}$ the relative position
vector pointing from particle $j$ to particle $i$, and $h$ the smoothing
length. Throughout this work we employ the quintic spline kernel,
\begin{equation}
W =\phi \begin{cases}
  (3-\bar{r})^5-6(2-\bar{r})^5+15(1-\bar{r})^5 & 0 \le \bar{r} < 1, \\
  (3-\bar{r})^5-6(2-\bar{r})^5 & 1 \le \bar{r} < 2, \\
  (3-\bar{r})^5 & 2 \le \bar{r} < 3, \\
  0 & \bar{r} \ge 3,
\end{cases}
\label{quintic}
\end{equation}
where $\bar{r} = r_{ij}/h$ and $r_{ij}=\left|\mathbf{r}_{ij}\right|$. The
normalization coefficient $\phi$ equals $1/(120\pi)$ in three dimensions.
The smoothing length is set to $h=\Delta x$, where $\Delta x$ is the
initial inter-particle spacing.

It is convenient to introduce an equivalent particle volume,
\begin{equation}
V_i = \frac{1}{\sum_{j}W_{ij}},
\label{equivalent_volume}
\end{equation}
which is consistent with $V_i = m_i/\rho_i$. In terms of this quantity,
the pressure-gradient term is discretized in the symmetric form
\begin{equation}
-\left( \frac{1}{\rho}\nabla p \right)_i
= - \frac{1}{m_i} \sum_{j}\left( V_{i}^{2}p_i + V_{j}^{2}p_j \right)
\frac{\partial W}{\partial r_{ij}}\,\mathbf{e}_{ij},
\label{pressure_sph}
\end{equation}
where $\mathbf{e}_{ij}=\mathbf{r}_{ij}/r_{ij}$ is the unit vector along
$\mathbf{r}_{ij}$, and the pressures $p_i$ and $p_j$ follow from $\rho_i$
and $\rho_j$ through Eq.~(\ref{state}). To account for the viscosity
contrast between the two phases, the viscous term is evaluated following
the multiphase formulation of Hu and
Adams~\cite{hu2006MultiphaseSPHMethod},
\begin{equation}
\left(\mu\nabla^{2}\mathbf{v}\right)_i
=\frac{1}{m_i} \sum_{j}\frac{2\mu_i \mu_j}{\mu_i + \mu_j}
\left( V_i^2 + V_j^2 \right)
\frac{\mathbf{v}_{ij}}{r_{ij}}\frac{\partial W}{\partial r_{ij}},
\label{visco_sph}
\end{equation}
in which $\mathbf{v}_{ij}=\mathbf{v}_{i}-\mathbf{v}_{j}$ is the relative
velocity of particles $i$ and $j$, and the harmonic mean of the
viscosities $2\mu_i\mu_j/(\mu_i+\mu_j)$ ensures a smooth treatment of the
viscosity jump across the interface.

Within this weakly compressible scheme, the density fluctuation is kept
below $\delta_\rho \le 0.5\%$ so as to approximate incompressible flow.
The artificial sound speed $c_{s}$ must be chosen large enough to
suppress these fluctuations. Following Morris
\textit{et al.}~\cite{morris2000simulating,morris1997ModelingLowReynolds},
$c_{s}^{2}$ is taken comparable to the largest of
\begin{equation}
\frac{U^2}{\delta_\rho},\quad \frac{\mu U}{\rho_{\mathrm{ref}} L \delta_\rho},\quad
\frac{FL}{\delta_\rho},
\label{cs_chosen}
\end{equation}
where $U$, $L$ and $F$ are the characteristic velocity, length and body
force, respectively. To minimize the pressure error,
$\rho_{\mathrm{ref}}$ is set as close as possible to, but strictly below,
the minimum density in the domain, thereby avoiding negative pressures.
The specific parameter settings follow our previous
work~\cite{wang2023dynamics}.

Time integration is carried out with the explicit velocity--Verlet
scheme, and the time step is chosen to satisfy the usual stability
constraints~\cite{morris2000simulating}. 
To maintain numerical stability with respect to the bending forces when using
explicit time integration, the capillary-wave CFL condition for a
droplet with surface tension in~\cite{morris2000simulating} 
is replaced by an analogous condition based on the
bending wave phase velocity. Since the bending energy is quadratic in the local
curvature, the surface tension $\sigma$ acts as an
effective tension $\sigma_{\rm eff}=\kappa_{b}{C_h}^2$, 
where $C_h\sim 2\pi/h$ is the characteristic curvature resolved
by the particle spacing $h$. Substituting $\sigma\to \kappa_{b}(2\pi/h)^2$ into
the condition for surface tension, yields the bending stability condition

\begin{equation}
\Delta t_b \le 0.25\left[\frac{\tilde\rho h^5}{8\pi^3 \kappa_{b}}\right]^{1/2}.
\label{eq:ds_bend}
\end{equation}

The remaining constraints are the standard ones for weakly compressible SPH,
arising from acoustic propagation at the artificial sound speed $c_s$, from the
local SPH particle acceleration $a$, and from viscous diffusion. Collecting these
together with Eq.~\eqref{eq:ds_bend}, the time step is advanced subject to
\begin{equation}
\Delta t=\min\left\{
\Delta t_b,\;
0.25\,\frac{h}{c_s},\;
0.25\min_j\left(\frac{h}{a_j}\right)^{1/2},\;
0.125\,\frac{\rho h^{2}}{\mu}
\right\}.
\label{eq:time_step}
\end{equation}
In all cases reported below the bending constraint is the most restrictive,
which reflects the stiffness introduced by the fourth-order dependence of the
bending force on the membrane curvature.

\subsubsection{Coupling the vesicle mesh with the SPH fluid}

\begin{figure}[htbp]
\centering
\includegraphics[width=0.6\columnwidth, keepaspectratio]{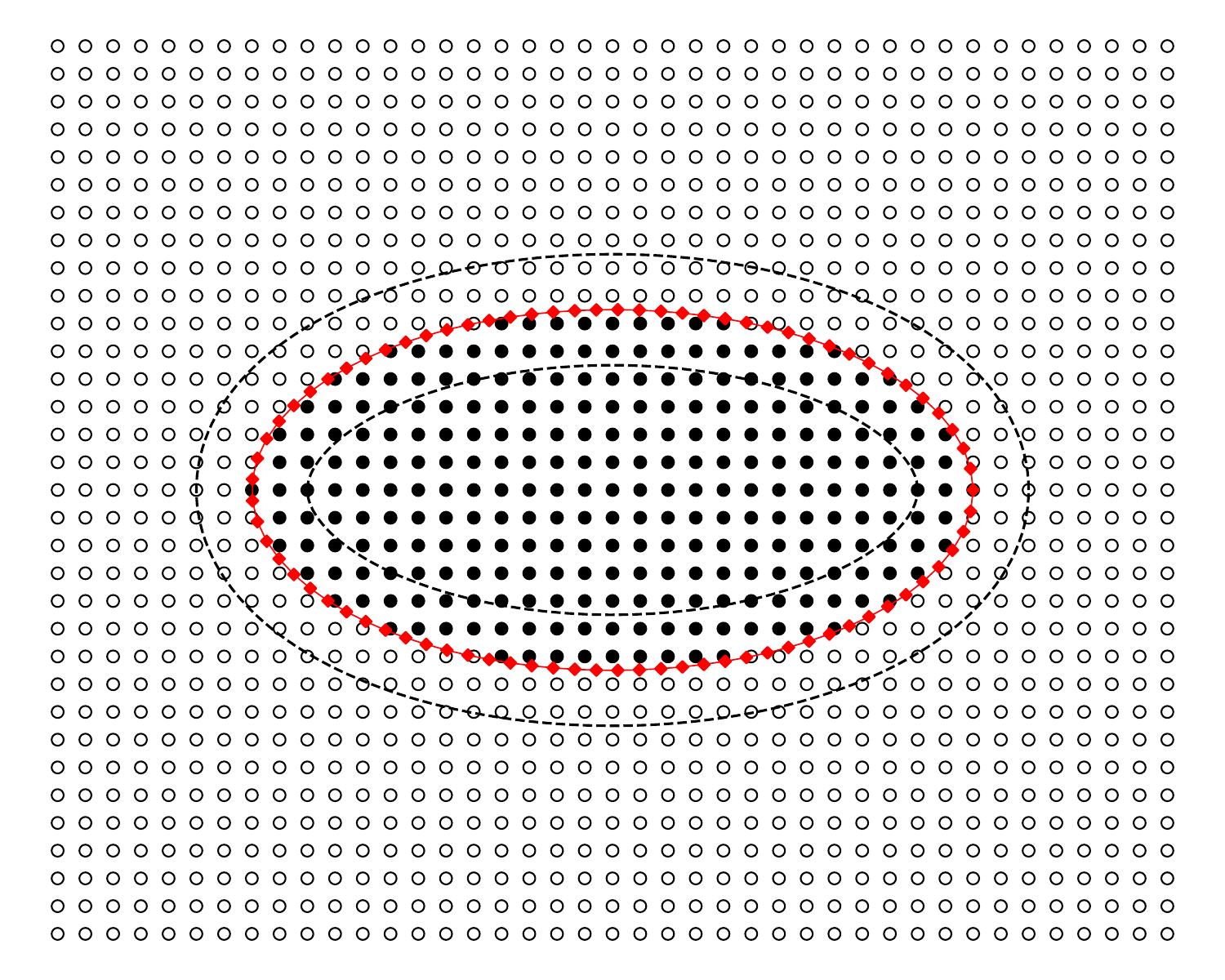}
\caption{\label{fig:sph_mesh_ibm} Two-dimensional schematic of the
coupling between the triangulated membrane and the SPH fluid. The red line
with diamonds represents the vesicle membrane. Hollow circles denote the
outer fluid particles and solid black circles the inner fluid particles.
The fluid particles lying between the dashed lines are those that exchange
both force and position information with the membrane mesh.}
\end{figure}

The membrane is coupled to the fluid through a new immersed-boundary
method, illustrated schematically in two dimensions in
Fig.~\ref{fig:sph_mesh_ibm}. The central idea is that the position of the
triangulated membrane is slaved to the interface between the inner and
outer fluids, while the mechanical forces evaluated on the membrane are
distributed back to the fluid particles located in a thin layer
surrounding the interface. In this way the membrane and the fluid
influence each other in a two-way fashion.

The interface is identified by means of a color function. At an arbitrary
position $\mathbf{x}$ it is defined as
\begin{equation}
c_\mathbf{x} = \frac{\sum_{i} c_{i} W_{\mathbf{x}i}}{\sum_{i}W_{\mathbf{x}i}},
\label{color_function}
\end{equation}
where $c_{i}$ is the color label of particle $i$ and $W_{\mathbf{x}i}$ is
the same kernel as in Eq.~(\ref{quintic}). Particles belonging to the
inner fluid are assigned $c_{i}=1$, and those belonging to the outer
fluid $c_{i}=0$, so that the interface corresponds to the iso-surface
$c_\mathbf{x}=0.5$. A driving force is applied to each membrane vertex to
push it towards this iso-surface,
\begin{equation}
\mathbf{F}_{d}^{\,j} = m_v\, r_c\, k_c^2 \left( 2 c_j - 1\right)\mathbf{n}_j,
\label{ibm_force}
\end{equation}
where $m_v$ is a virtual mass assigned to the vertex, $r_c = 3h$ is the
cutoff radius of the kernel used in Eq.~(\ref{color_function}), and $k_c$
is a control coefficient that allows the membrane to be held at 
the interface accurately and stably. We set the virtual mass $m_v$ 
equal to the mass of a single SPH fluid particle, and the coupling
stiffness $k_c$ equal to the numerical speed of sound $c_s$. 
This choice is demonstrated, through both theoretical scaling 
analysis and extensive numerical tests, to yield stable and accurate 
interface tracking, while also enabling these new parameters to 
scale naturally with the original SPH system. 
Here $c_j$ is the color value evaluated
at the position of vertex $j$ through Eq.~(\ref{color_function}), and
$\mathbf{n}_j$ is the outward normal at vertex $j$, obtained as the
area-weighted average of the normals of the incident triangles,
\begin{equation}
\mathbf{n}_j= \frac{\sum_i \mathbf{n}_i A_{i}^t}{\sum_i A_{i}^t}.
\label{vertex_normal}
\end{equation}
The sign factor $(2c_j-1)$ ensures that the driving force always points
towards the $c_\mathbf{x}=0.5$ surface, restoring any vertex that has
drifted to either side of the interface.

Simultaneously, each vertex experiences a viscous force that dissipates
the kinetic energy generated by the driving force and thereby stabilizes
the coupling. This force is constructed from the viscosity of the
neighboring fluid,
\begin{equation}
\mathbf{F}_{v}^{\,j}=\sum_{i} \mu_i\left( V_i^2 + V_j^2 \right)
\frac{\mathbf{v}_{ji}}{r_{ij}}\frac{\partial W}{\partial r_{ij}}.
\label{ibm_viscous_force}
\end{equation}
In addition to its stabilizing role, this viscous coupling makes the
membrane follow the tangential motion of the surrounding fluid, which is
essential for reproducing the tank-treading behavior of the membrane.
Together, the driving force~(\ref{ibm_force}) and the viscous
force~(\ref{ibm_viscous_force}) determine the motion of the mesh vertices
and constitute the total force exerted on them.

The forces evaluated on the membrane vertices are transferred to the
neighboring fluid particles through an SPH-kernel interpolation,
\begin{equation}
\mathbf{F}_{i}^{b} = \sum_{j} \mathbf{F}_{j}\,
\frac{W_{ij}}{\sum_{k}W_{kj}},
\label{ibm_distribute}
\end{equation}
where $\mathbf{F}_{i}^{b}$ is the force distributed to fluid particle $i$,
$\mathbf{F}_{j}$ is the force computed on membrane vertex $j$, and the
index $k$ runs over the fluid particles neighboring vertex $j$. The
kernels $W_{ij}$ and $W_{kj}$ are again those of Eq.~(\ref{quintic}), and
the normalization by $\sum_{k}W_{kj}$ guarantees that the total force is
conserved during the transfer.

Because the fluid particles prescribe the \emph{position} of the membrane
rather than its velocity, the mesh moves together with the interface
between the two fluids. Consequently, no additional non-physical
reflection or bounce-back conditions are required to prevent fluid
particles from crossing the membrane, which is a distinctive advantage of
the present coupling strategy.

\subsubsection{Area and volume constraints}

Since the number of fluid particles enclosed by the vesicle is fixed, the
volume of the vesicle is conserved intrinsically through the weak
compressibility of the SPH fluid, and no explicit volume constraint is
needed. Only the membrane area has to be constrained. Treating the
membrane as a weakly compressible surface, we introduce a global and a
local area-constraint energy that penalize, respectively, the variation
of the total area and of each individual triangle,
\begin{equation}
E^{A}=\kappa_{a g}\frac{\left(A-A_{0}\right)^{2}}{A_{0}}
+ \kappa_{a l}\sum_{i=1}^{N_{t}}
\frac{\left(A_{i}^{t}-A_{0}^{t}\right)^{2}}{A_{0}^{t}},
\label{area_energy}
\end{equation}
where $\kappa_{a g}$ and $\kappa_{a l}$ are the global and local
area-constraint moduli, and $A_{0}$ and $A_{0}^{t}$ are the target areas
of the whole vesicle and of a single triangle, respectively. We use
$\kappa_{a g} = 2000\kappa_{b} / R^2$ and $\kappa_{a l} = 1000\kappa_{b} / R^2$ in our
simulations. The corresponding constraint force is obtained from the
negative gradient of this energy,
\begin{equation}
\mathbf{F}^{A}=-\frac{\partial E^{A}}{\partial \mathbf{x}}.
\label{area_constraint_force}
\end{equation}
Finally, a triangle equiangulation procedure~\cite{brakke1992surface}, 
also called the bond-flipping scheme, is
periodically applied to preserve the quality and stability of the mesh;
as a by-product, this operation also permits an in-plane flow of the
membrane, consistent with the fluid nature of the lipid bilayer.

\section{\label{sec:numerical results}Numerical results}

\begin{figure}[htbp]
\centering
\includegraphics[width=0.6\columnwidth, keepaspectratio]{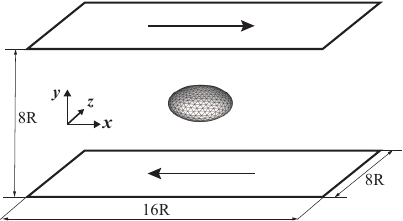}
\caption{Sketch of the simulation set-up. A single vesicle is initially
placed at the centre of a rectangular box of dimensions
$16R \times 8R \times 8R$ in the $x$, $y$ and $z$ directions, where $R$
is the radius of a sphere having the same surface area as the vesicle.
The shear flow is driven by the two solid walls at $y = \pm 4R$, which
move in opposite directions along $x$ with equal speed, as indicated by
the arrows, giving a nominal shear rate $\dot\gamma$. Periodic boundary
conditions are imposed in the $x$ and $z$ directions. The membrane is
discretised into triangular elements and the fluid is represented
by SPH particles (only the membrane
mesh is shown for clarity).\label{fig:initial_box_shear}}
\end{figure}

The present method is developed to investigate the dynamics of vesicles
with complex and non-trivial shapes under shear flow. Unless stated
otherwise, the target area of the vesicle is $A_0=4 \pi R^2$ with $R=1$.
The vesicle is placed at the centre of a rectangular box of size
$L_x = 16R$, $L_y = 8R$, and $L_z = 8R$, filled with fluid, as shown in
Fig.~\ref{fig:initial_box_shear}. Solid plates are located at $y=-4R$ and
$y=4R$, while the remaining two directions are treated as periodic. A
no-slip condition is imposed on the plates, which move in opposite
directions with speed $U$ to impose a shear rate $\dot{\gamma}$ on the
enclosed fluid. In the present work, we restrict our attention to the
case where the fluids inside and outside the vesicle share the same
density and viscosity, i.e.\ $\mu_i = \mu_j = \mu$. In analogy with a
droplet, a capillary number based on the bending modulus is introduced to
measure the ratio of viscous to bending forces,
$Ca = \mu \dot{\gamma} R^{3} / \kappa_b$, where $\dot{\gamma} = 2U/L_y$ is
the shear rate. The Reynolds number is defined as
$Re = \rho \dot{\gamma} R^{2} / \mu$. In order to mimic the
low-Reynolds-number conditions typical of microchannel flows while
keeping the computational cost affordable, all simulations are carried
out at $Re = 0.1$. The remaining parameters are fixed at $\rho = 1000$ and
$\dot{\gamma} = 1 \times 10^{-5}$, and the values of $\mu$ and $\kappa_b$
are tuned to realize the prescribed $Re$ and $Ca$.

In this work, the triangulated polyhedron is constructed from an initial 
regular icosahedron by iteratively subdividing each triangle into four 
smaller triangles, projecting onto the target vesicle geometry, and 
equalizing triangle areas. The vesicle resolution thus scales as 
20, 80, 320, 1280, 5120, ... triangles per vesicle. 
As shown by Bian \textit{et al.}~\cite{bian2020bending}, a triangulated
mesh of 1280 triangles is already sufficient to reproduce the various
equilibrium shapes of vesicles. Accordingly, each vesicle is discretized
with 1280 triangular elements in all simulations reported below. The SPH
particles are initialized on a cubic lattice with an inter-particle
spacing of $\Delta x = 0.2\,R$, comparable to the edge length of the
membrane triangles, which provides a well-balanced resolution that
exploits the strengths of both discretizations. For the convergence study
presented in the validation section, an additional set of
higher-resolution simulations was performed, in which each vesicle is
discretized with 5120 triangles and the particle spacing is refined to
$\Delta x = 0.1\,R$. It should be emphasized that the particle resolution
need not match that of the membrane mesh; the membrane mesh, however, must
be resolved more finely than the fluid, so as to prevent the particles
from penetrating the membrane.

\subsection{\label{sec:validation}Validation}

To assess the reliability of the proposed method, we first reproduce a
series of static equilibrium shapes of vesicles and the dynamics of a
simple vesicle in shear flow, all of which have been documented in
previous studies and therefore serve as well-established references.

\subsubsection{Static vesicle shapes}

A rich variety of vesicle shapes has been reported in experiments,
including prolates, biconcave discocytes, dumbbells, stomatocytes and
triangular oblates. Such morphological diversity can be rationalized
within the bending models implemented in our solver. Bian
\textit{et al.}~\cite{bian2020bending} computed the equilibrium shapes of
vesicles over a wide parameter range using purely membrane-based meshes.
Here we select a representative subset of their cases to validate our
coupled SPH--mesh method: the aim is to demonstrate that, once the
surrounding fluid is introduced, the equilibrium shapes predicted by our
method coincide with those obtained from the purely mesh-based model.

For the static-shape tests the shear rate is set to $\dot{\gamma}=0$, so
that the vesicle relaxes in a quiescent fluid. Two families of initial
configurations are considered, oblate and prolate, both with initial area
$A_0=4\pi R^2$ and with an initial volume equal to the target volume. The
reduced volume, defined as $V_r = V/(\tfrac{4}{3}\pi R^3)$, measures the
vesicle volume relative to that of a sphere of the same area and serves as
the principal control parameter.
Starting from these initial configurations, the system is evolved until the
total energy reaches a plateau and the shape no longer changes; the
resulting configuration is taken as the equilibrium shape (see
Appendix~\ref{app:particles} for the corresponding particle distribution).

\begin{figure*}[tbp]
\centering
\includegraphics[width=170mm]{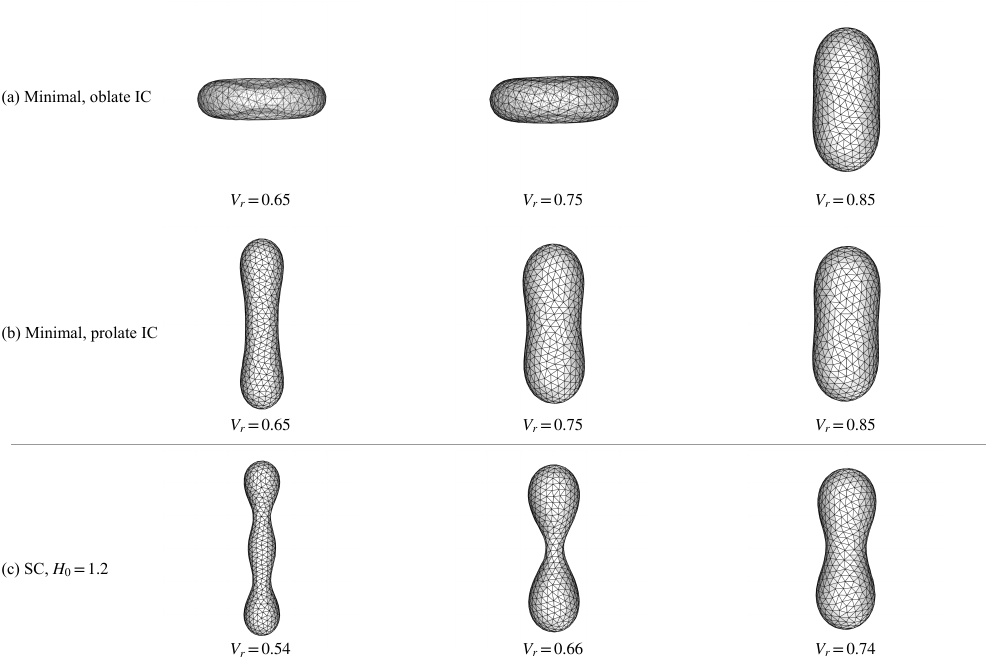}
\caption{Equilibrium shapes without and with local spontaneous curvature.
(a),(b) Minimal model ($H_0=0$, $\alpha=0$) started from oblate and prolate
initial conditions, respectively. (c) SC model with $H_0=1.2$. All shapes
are rendered at the same scale and viewed along the same
direction.\label{fig:steady_minimal_sc}}
\end{figure*}

We first contrast the minimal model with the SC model, which isolates the
effect of a local spontaneous curvature. The minimal model is recovered by
setting $H_0=0$ and $\alpha=0$ in Eq.~(\ref{total_energy_integral}); the
resulting shapes are shown in Figs.~\ref{fig:steady_minimal_sc}(a) and
\ref{fig:steady_minimal_sc}(b). At $V_r=0.65$ and $V_r=0.75$ the final state
retains a memory of the initial condition: the oblate initialization
relaxes into a biconcave discocyte, whereas the prolate initialization
settles on a prolate shape. At $V_r=0.85$ this dependence disappears and
both initializations converge to the same equilibrium. Such bistability is
a known feature of the Helfrich energy landscape at low reduced volumes,
where the discocyte and prolate branches are separated by an energy
barrier~\cite{seifert1991shape}. Introducing a finite spontaneous curvature
$H_0=1.2$ changes the picture qualitatively
(Fig.~\ref{fig:steady_minimal_sc}(c)): the membrane now favours a preferred
local curvature, and narrow necks develop, giving rise to two-lobed and
multi-lobed configurations that are inaccessible to the minimal model. All
simulations in this and the following comparison are initialized from an
oblate ellipsoid.

\begin{figure*}[tbp]
\centering
\includegraphics[width=170mm]{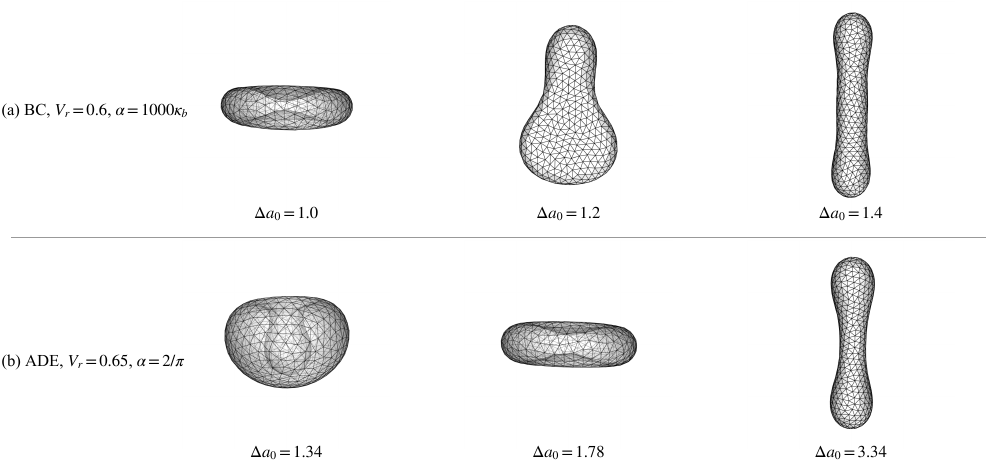}
\caption{Equilibrium shapes under a strong and a weak non-local
area-difference constraint. (a) BC model with $V_r=0.6$ and
$\alpha=1000$, which effectively enforces the hard constraint
$\Delta A=\Delta A_0$. (b) ADE model with $V_r=0.65$ and $\alpha=2/\pi$,
for which the bending and area-difference energies are comparable. All
shapes are rendered at the same scale.\label{fig:steady_nonlocal}}
\end{figure*}

We next compare the BC and ADE models, which differ in the strength with
which the non-local area difference is constrained. It is convenient to
normalize the area differences as $\Delta a_0=\Delta A_0/(4\pi R D)$ and
$\Delta a=\Delta A/(4\pi R D)$. In the BC model we set
$\alpha=1000$, large enough to effectively enforce the hard
constraint $\Delta A=\Delta A_0$; the corresponding shapes are given in
Fig.~\ref{fig:steady_nonlocal}(a). In contrast to the two local models,
the BC model produces markedly asymmetric, pear-like shapes, reflecting
the strong coupling between the imposed area difference and the membrane
geometry. When $\alpha$ is reduced so that the bending and area-difference
energies become comparable, the two contributions compete and the ADE
regime is recovered. Setting $\alpha=2/\pi$
(Fig.~\ref{fig:steady_nonlocal}(b)), the constraint is only soft, and an
increasing preferred area difference $\Delta a_0$ drives the vesicle
through a sequence of increasingly elongated morphologies rather than
locking it onto a single asymmetric shape.

For every parameter set considered above, the steady shapes produced by
our solver have a direct counterpart in the purely mesh-based results of
Bian \textit{et al.}~\cite{bian2020bending}: the same branches are
selected, the same neck-forming and pear-like morphologies appear, and the
transitions occur at the same values of $V_r$ and $\Delta a_0$. This
agreement across all four bending models confirms that coupling the
vesicle mesh to the SPH fluid does not compromise the accuracy of the
predicted equilibrium shapes. Unlike the purely mesh-based approach,
however, our method simultaneously resolves the surrounding fluid,
including the internal flow and the flow around the vesicle, which is a
prerequisite for the dynamic simulations presented in the following
sections.

\subsubsection{Inclination and revolution frequency of a vesicle in shear flow by the minimal model}

To validate our numerical framework, we examined the steady-state
tank-treading behavior of a single vesicle subjected to a linear shear
flow using the minimal model over a range of reduced volumes
$V_r \in [0.5, 1.0]$, considering two capillary numbers ($Ca=1$ and
$Ca=10$). After the vesicle reaches its equilibrium shape, the top and
bottom plates are moved in opposite directions to generate the shear
flow. The vesicle then undergoes a tank-treading motion. We calculated
the inclination angle and revolution frequency of the vesicle after it
had reached a dynamically stable state under shear flow. The inclination
angle is defined as the angle between the longest axis of the vesicle and
the $x$ direction. The revolution frequency was determined from the
average of the time intervals between the two most recent sign changes of
the $y$ coordinate for each vertex on the vesicle mesh.

\begin{figure*}[tbp]
    \centering
    \includegraphics[width=0.95\textwidth, keepaspectratio]{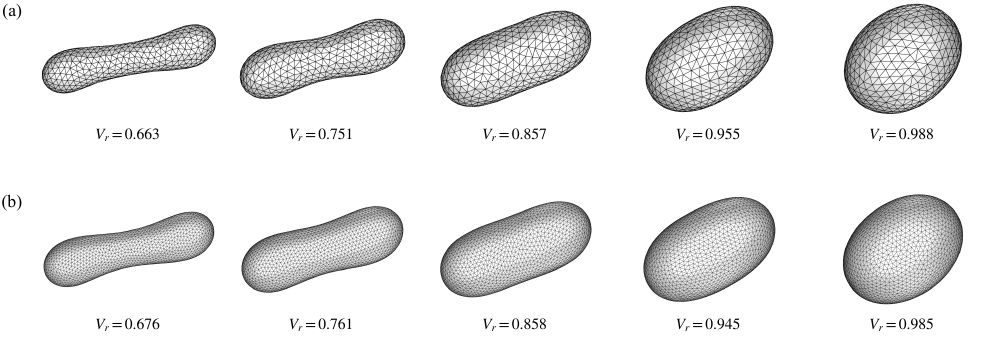}
    \caption{\label{fig:shape_compare_high} Comparison of the dynamic
    equilibrium shapes of vesicles in shear flow obtained with two mesh
    resolutions using the minimal model at $Ca = 1$. (a) Top row: the standard
    resolution with $1280$ triangular elements and a particle spacing of
    $\Delta x = 0.2R$. (b) Bottom row: the high resolution with $5120$
    triangular elements and a particle spacing of $\Delta x = 0.1R$. From
    left to right, the reduced volume increases progressively.}
\end{figure*}

To assess the sensitivity of the present method to the spatial
resolution, we performed an additional set of simulations at a higher
resolution, in which the vesicle membrane is discretized with $5120$
triangular elements and the fluid is resolved with a particle spacing of
$\Delta x = 0.1R$, as opposed to the $1280$ elements and
$\Delta x = 0.2R$ employed in all the other cases. Figure~\ref{fig:shape_compare_high}
compares the dynamic equilibrium shapes obtained with the two
resolutions across a range of reduced volumes for the minimal model at
$Ca = 1$. The morphologies at comparable $V_r$ are virtually
indistinguishable, indicating that the equilibrium shape is well
converged already at the standard resolution.

\begin{figure*}[tbp]
    \centering
    \begin{minipage}[b]{0.48\textwidth}
        \centering
        \includegraphics[width=\linewidth]{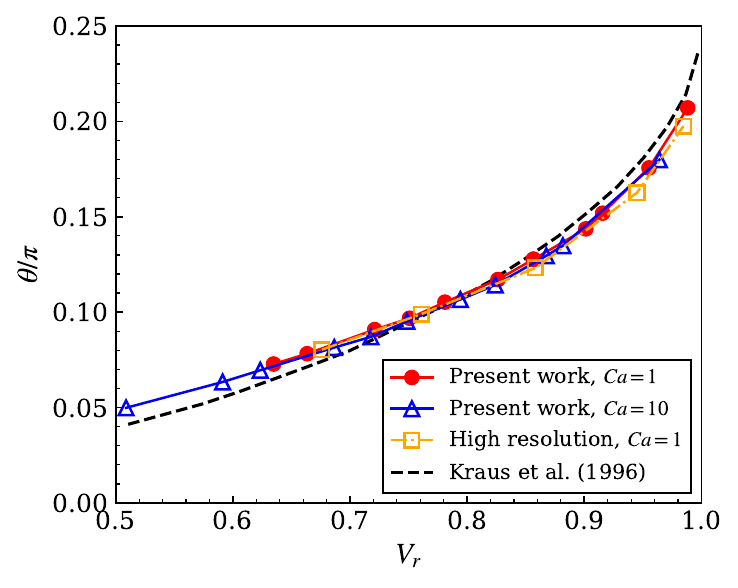}\\
        (a)
    \end{minipage}
    \hfill
    \begin{minipage}[b]{0.48\textwidth}
        \centering
        \includegraphics[width=\linewidth]{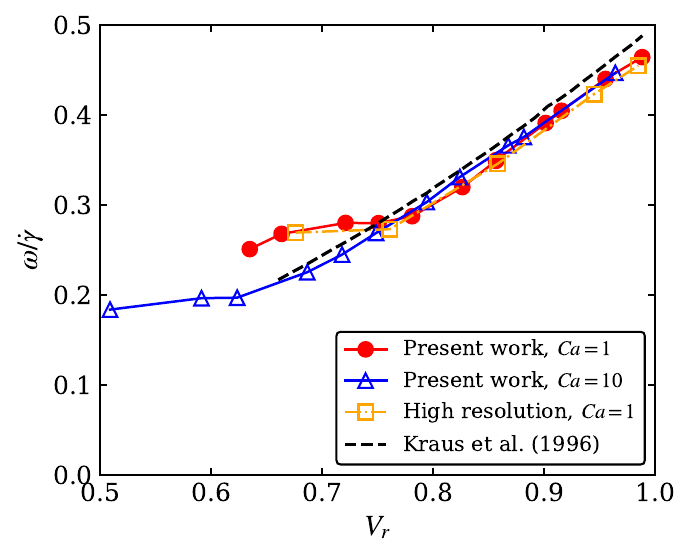}\\
        (b)
    \end{minipage}
    \caption{Inclination (a) and revolution frequency (b) of the vesicle
    in shear flow versus the reduced volume using the minimal model under
    $Ca = 1$ and $Ca = 10$, compared with the theoretical results of Kraus
    \textit{et al.}~\cite{kraus1996fluid}. The open squares denote the
    high-resolution results ($5120$ elements, $\Delta x = 0.1R$) at
    $Ca = 1$, which are in good agreement with the standard-resolution
    results ($1280$ elements, $\Delta x = 0.2R$).}
    \label{fig:compare_kraus1996}
\end{figure*}

A more quantitative comparison is presented in
Fig.~\ref{fig:compare_kraus1996}, where the high-resolution results for
the inclination angle and revolution frequency are superimposed on the
standard-resolution data. The two data sets are in good agreement over
the entire range of reduced volumes, with only minor deviations that
remain well within the scatter of the measurements. This convergence
study confirms that the standard resolution of $1280$ elements and
$\Delta x = 0.2R$ is sufficient to capture both the equilibrium
morphology and the dynamical response of the vesicle, while keeping the
computational cost affordable. Consequently, this resolution is adopted
throughout the remainder of this work.

\begin{figure*}[tbp]
\centering
\includegraphics[width=\textwidth]{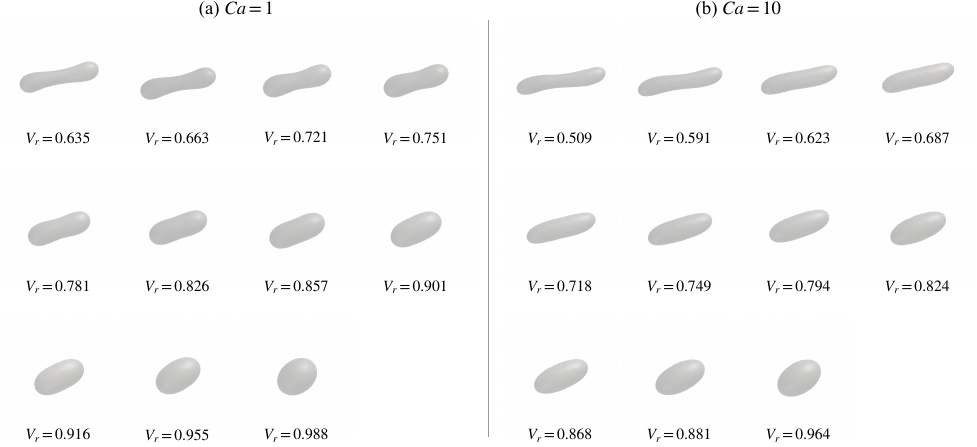}
\caption{Dynamic equilibrium shapes of a vesicle in shear flow obtained
with the minimal model at (a) $Ca = 1$ and (b) $Ca = 10$. The reduced
volume $V_r$ is indicated below each shape and increases from left to
right and from top to bottom. All shapes are viewed in the shear plane,
with the flow directed along the horizontal axis, and are rendered at the
same scale. At low $V_r$ and $Ca = 1$ the vesicle develops an oblong
shape with a central indentation, which deviates from the ellipsoidal
geometry assumed in the theory of Kraus \textit{et
al.}~\cite{kraus1996fluid}.\label{fig:shear_minimal}}
\end{figure*}

Figure~\ref{fig:compare_kraus1996} also shows the results of our simulations
compared with the theory of Kraus \textit{et
al.}~\cite{kraus1996fluid}. The inclination and revolution frequency of
the vesicle vary with the reduced volume, consistent with theoretical
predictions at large reduced volume. We observe some discrepancies: when
$Ca = 1$ and $V_r < 0.75$, our results deviate from the theoretical
predictions. By analyzing the dynamic equilibrium shape of the vesicle in
our simulations, as shown in Fig.~\ref{fig:shear_minimal}, 
we found that the vesicle exhibits an
oblong shape with a central indentation. This morphology arises at small
$Ca$, where the bending force can counteract the shear force, leading to
a shape closer to a biconcave one, deviating from the ellipsoidal shape.
Notably, the theoretical framework assumes that vesicles maintain an
ellipsoidal shape. This accounts for the observed discrepancy between the
simulation and the theoretical prediction.

\subsection{Simulation of a vesicle in shear flow with different bending models}

We next explore the dynamics of vesicle behavior under shear flow in the
various models. For the SC model, we compare the inclination angle and
revolution frequency with those of the minimal model. For the BC and ADE
models, we focus primarily on identifying special vesicle shapes and
behaviors.

\subsubsection{SC model: inclination and revolution frequency}

\begin{figure*}[tbp]
    \centering
    \begin{minipage}[b]{0.48\textwidth}
        \centering
        \includegraphics[width=\linewidth]{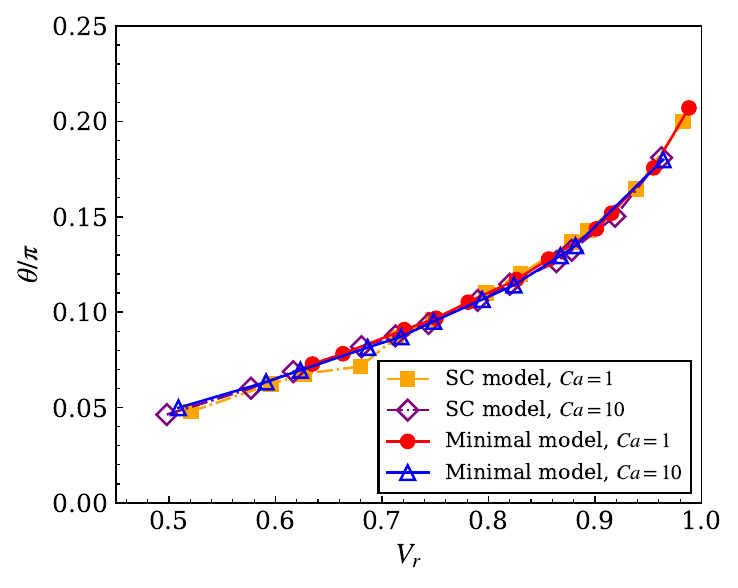}\\
        (a)
    \end{minipage}
    \hfill
    \begin{minipage}[b]{0.48\textwidth}
        \centering
        \includegraphics[width=\linewidth]{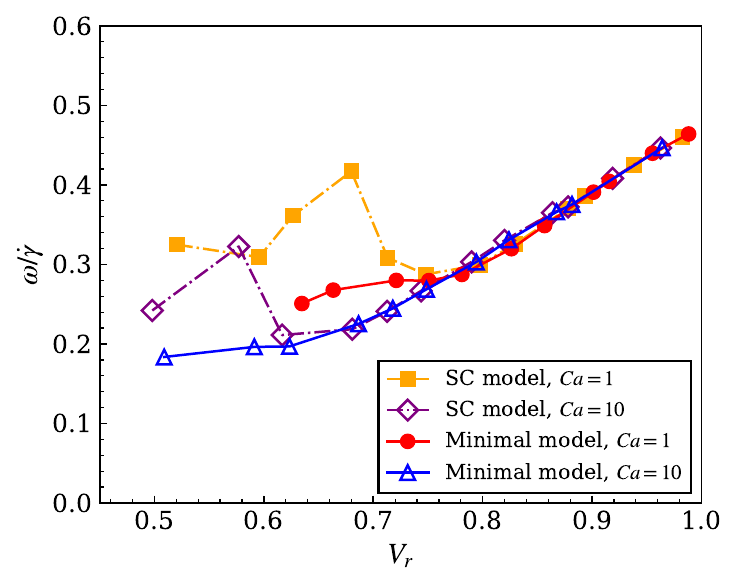}\\
        (b)
    \end{minipage}
    \caption{(a) Inclination angle $\theta/\pi$ and (b) normalised revolution
frequency $\omega/\dot\gamma$ of a vesicle in shear flow as a function of
the reduced volume $V_r$, obtained with the SC model ($H_0 = 1.2$) and
compared with the minimal model, for $Ca = 1$ and $Ca = 10$. Filled
symbols denote the minimal model and open symbols the SC model; circles
and squares correspond to $Ca = 1$, triangles and diamonds to $Ca = 10$.
The inclination angle is essentially independent of the bending model
over the whole range of $V_r$. The revolution frequencies of the two
models coincide for $V_r \gtrsim 0.8$, but differ markedly at smaller
reduced volumes, where the SC model yields higher and strongly
non-monotonic frequencies associated with the deeply necked,
dumbbell-like shapes shown in Fig.~\ref{fig:shear_sc}.
\label{fig:theta_omega_sc}}
\end{figure*}

We investigate the influence of a non-zero spontaneous curvature on the
tank-treading dynamics of a vesicle in shear flow. The SC model is
implemented with a spontaneous curvature $H_0 = 1.2$, and the resulting
inclination angle and revolution frequency are compared with those of the
minimal model in Fig.~\ref{fig:theta_omega_sc}.

As shown in Fig.~\ref{fig:theta_omega_sc}(a), the inclination angle obtained from
the SC model is in close agreement with that of the minimal model over
the entire range of reduced volumes for both capillary numbers. The angle
increases monotonically with $V_r$ and is essentially insensitive to the
presence of spontaneous curvature. This indicates that the orientation of
the vesicle is governed primarily by its overall geometric anisotropy,
which is set by the reduced volume, rather than by the local details of
the bending model.

In contrast, the revolution frequency exhibits a pronounced
model-dependent behavior at small reduced volumes, as illustrated in
Fig.~\ref{fig:theta_omega_sc}(b). For $V_r \gtrsim 0.8$, the SC model and the
minimal model yield nearly identical frequencies that grow monotonically
with $V_r$, reflecting the near-ellipsoidal shapes shared by both models
in this regime. However, for $V_r \lesssim 0.75$, the SC model produces
significantly higher revolution frequencies than the minimal model, and
the frequency curves become strongly non-monotonic: at $Ca = 1$ a
distinct local maximum emerges near $V_r \approx 0.68$, while at $Ca = 10$
the frequency first decreases to a local minimum near $V_r \approx 0.58$
before rising again at smaller $V_r$.

\begin{figure*}[tbp]
\centering
\includegraphics[width=\textwidth]{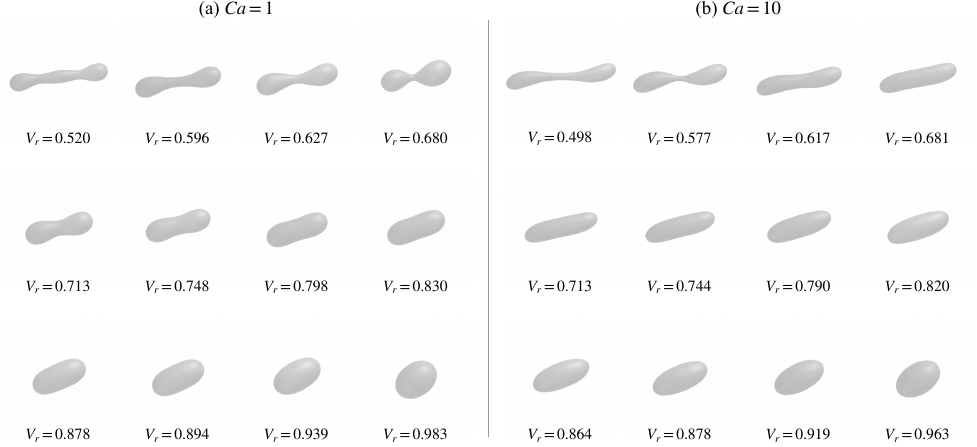}
\caption{Dynamic equilibrium shapes of a vesicle in shear flow predicted
by the SC model with $H_0 = 1.2$, at (a) $Ca = 1$ and (b) $Ca = 10$. The
reduced volume $V_r$ is indicated below each shape and increases from
left to right and from top to bottom; the viewing direction and the
rendering scale are the same as in Fig.~\ref{fig:shear_minimal}. As $V_r$
decreases, the excess area is accommodated by an increasingly deep
central neck, and at $Ca = 10$ the vesicle approaches a pinch-off
configuration at $V_r \approx 0.577$.\label{fig:shear_sc}}
\end{figure*}

\begin{figure*}[tbp]
    \centering
    \begin{minipage}[b]{0.45\textwidth}
        \centering
        \includegraphics[width=\linewidth]{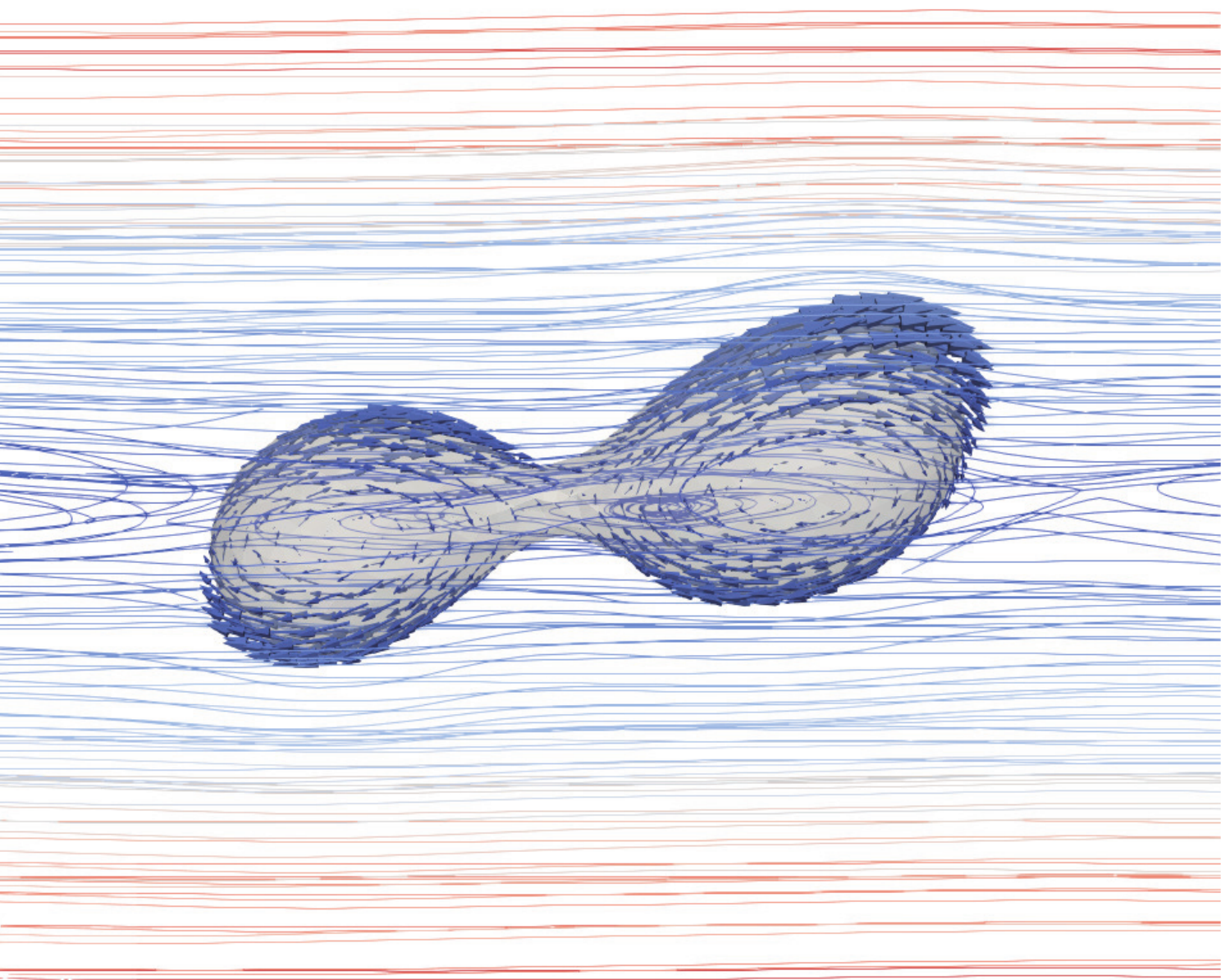}\\
        (a)
    \end{minipage}
    \hfill
    \begin{minipage}[b]{0.45\textwidth}
        \centering
        \includegraphics[width=\linewidth]{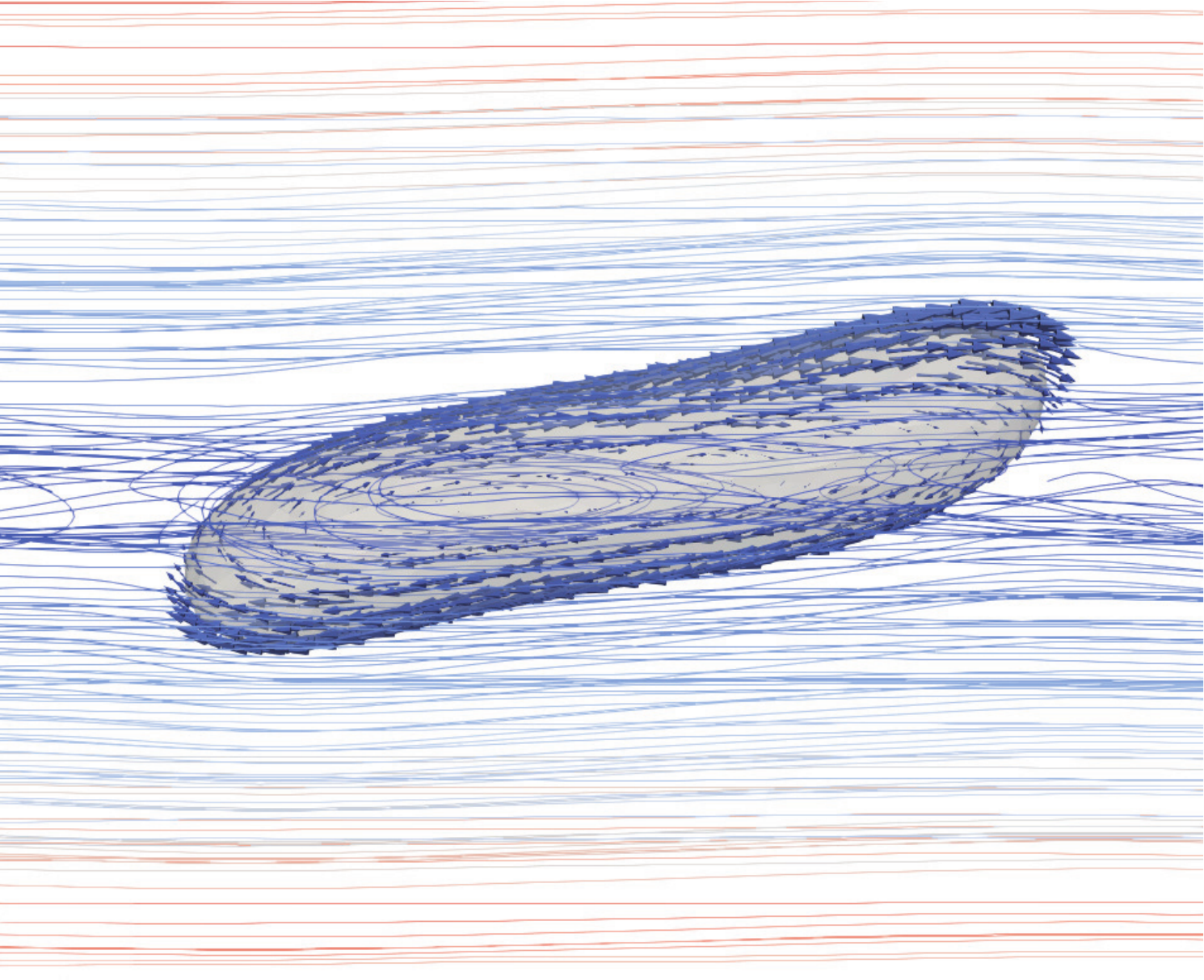}\\
        (b)
    \end{minipage}
    \caption{Velocity distribution on the membrane (arrows) and
    streamlines of the surrounding fluid for the SC model. (a)
    $V_r = 0.680$, $Ca = 1$: the vesicle forms a dumbbell shape and the
    membrane velocity circulates predominantly within each individual
    lobe. (b) $V_r = 0.681$, $Ca = 10$: the vesicle is stretched into an
    elongated ellipsoid exhibiting a global tank-treading circulation.}
    \label{fig:stream_sc_v68}
\end{figure*}

These deviations can be explained by examining the dynamic equilibrium
shapes of the SC-model vesicle shown in Fig.~\ref{fig:shear_sc}. 
Because the spontaneous curvature
energetically favors regions of high local curvature, the excess area
available at small $V_r$ is accommodated by the formation of a deeply
constricted, dumbbell-like morphology with a pronounced central neck.
This necking becomes increasingly severe as the reduced volume decreases,
and at $Ca = 10$ the vesicle approaches a pinch-off configuration near
$V_r \approx 0.577$, almost separating into two lobes. Such non-convex,
dumbbell shapes differ markedly from the smooth oblong shapes obtained
with the minimal model (Fig.~\ref{fig:shear_minimal}). 
The concentration of membrane area
within the two lobes alters the tangential circulation of the membrane.
By examining the surface velocity field of vesicles with reduced volume
of about $0.68$ shown in Fig.~\ref{fig:stream_sc_v68}, we find that the
flow pattern depends strongly on the capillary number. At low $Ca$
(Fig.~\ref{fig:stream_sc_v68}(a)), the vesicle adopts a deeply necked,
dumbbell shape, and a considerable fraction of the mesh vertices
(representing part of the lipid molecules of the membrane) circulate
exclusively within a single lobe. This localization markedly reduces
their effective radii of circulation and thereby increases the average
angular velocity, which accounts for the enhanced revolution frequency of
the SC model at small reduced volumes. At high $Ca$
(Fig.~\ref{fig:stream_sc_v68}(b)), by contrast, the strong flow
stretches the vesicle into a smooth, elongated ellipsoid without a neck,
and the membrane recovers a global tank-treading circulation. As $V_r$
increases toward unity, the spontaneous-curvature constraint is more
easily satisfied by a smooth quasi-spherical shape, the neck disappears,
and the SC model results converge to those of the minimal model.

Overall, these results demonstrate that while the spontaneous curvature
has little effect on the steady orientation of the vesicle, it strongly
modifies the rotational dynamics at low reduced volumes through the
emergence of deeply necked, dumbbell-shaped morphologies that are absent
in the minimal model.

\subsubsection{BC model: asymmetric single-dimpled shapes}

In addition to the minimal and SC models, our framework also incorporates
the BC and ADE models. Unlike the local spontaneous curvature, these
models impose a non-local constraint on the area difference between the
two leaflets of the bilayer, which can stabilize morphologies that are
inaccessible to the simpler models. Based on the preceding analysis of
the revolution frequency of the SC model at small reduced volumes, we
have learned that, for complex morphologies, the membrane surface flow
can develop localized internal circulation confined to small regions,
which diminishes the physical significance of averaging the revolution
frequency over the entire vesicle. Therefore, for the BC and ADE models,
we no longer focus on the inclination angle and the revolution frequency
as output parameters, but instead concentrate on identifying special
vesicle morphologies.

\begin{figure}[htbp]
\centering
\includegraphics[width=0.6\columnwidth]{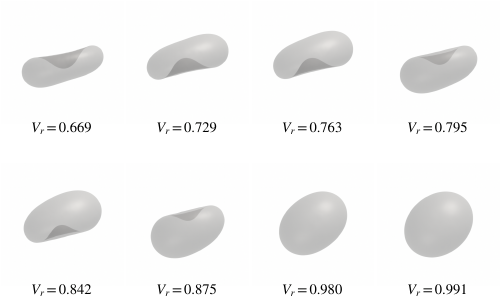}
\caption{Dynamic equilibrium shapes of a vesicle in shear flow obtained
with the BC model with $\alpha = 1000$ for a 
fixed reduced area difference $\Delta a_0 = 1$ at
$Ca = 1$. The reduced volume $V_r$ is indicated below each shape and
increases from left to right and from top to bottom. The membrane is
rendered semi-transparently so that the central indentation is visible.
Near-spherical vesicles remain smooth and convex;
upon deflation the vesicle first develops a central groove and adopts an
oblate, biconcave-like profile, and at the lowest reduced volumes it
elongates into an asymmetric slipper-like shape with a pronounced
off-center dimple. All shapes are viewed in the shear plane, with the
flow directed along the horizontal axis.\label{fig:shear_bc}}
\end{figure}

Figure~\ref{fig:shear_bc} presents the dynamic equilibrium shapes
obtained with the BC model for a fixed area difference
($\Delta a_0 = 1$) at $Ca = 1$, across a range of reduced volumes. A
clear morphological transition is observed as $V_r$ varies. For
near-spherical vesicles ($V_r \gtrsim 0.95$), the shape remains a smooth,
convex ellipsoid that is indistinguishable from those produced by the
other models, since the limited excess area leaves little freedom for
shape deformation.

As the reduced volume decreases, the additional excess area is
accommodated through the development of a central indentation: the vesicle
progressively flattens and adopts an oblate, biconcave-like profile
(e.g., $V_r = 0.842$ and $V_r = 0.875$), where the semi-transparent
rendering reveals the concave depression on the membrane surface. With
further deflation ($V_r \lesssim 0.76$), the vesicle elongates along the
flow direction while retaining the central groove, evolving into a
characteristic asymmetric, slipper-like morphology with a pronounced
off-center dimple. At the smallest reduced volumes investigated
($V_r = 0.669$), the shape becomes a strongly elongated, twisted
configuration in which the indentation is markedly skewed relative to the
long axis.

These morphologies differ qualitatively from the symmetric, dumbbell-like
shapes generated by the SC model at comparable reduced volumes. The
non-local area-difference constraint of the BC model penalizes global
asymmetries in leaflet area and thereby favors laterally extended, grooved
shapes rather than the deeply necked configurations driven by the local
spontaneous curvature. This demonstrates the capability of our coupled
SPH--mesh framework to capture the distinct shape selection arising from
non-local bending energetics, and highlights the richer morphological
repertoire accessible once the bilayer architecture is explicitly
accounted for.

\subsubsection{ADE model: a potential approach for generating vesicle-in-vesicle structures}

\begin{figure}[htbp]
\centering
\includegraphics[width=0.6\columnwidth]{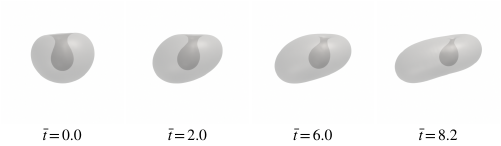}
\caption{Shape evolution of a stomatocyte vesicle predicted by the ADE
model under shear flow, for $V_r = 0.65$, $\Delta a_0 = 1.34$, $H_0 = 0$
and $Ca = 1$. Time is normalised as $\bar t = \dot\gamma\,t$ and
increases from left to right; the flow is directed along the horizontal
axis. The membrane is rendered semi-transparently so that the engulfed
inner cavity is visible. The vesicle tilts and elongates along the flow
while the inner cavity is drawn into a slender pouch. The last snapshot
($\bar t = 8.2$) is the frame immediately before the simulation
terminates, in which the neck connecting the inner and outer membranes
has thinned to the verge of rupture.\label{fig:shear_ade}}
\end{figure}

For the ADE model, our primary interest lies in the dynamical behavior of
the stomatocyte vesicle, corresponding to the first morphology of ADE model in
Fig.~\ref{fig:steady_nonlocal}(b), when subjected to shear flow. We adopt the
parameters $Ca = 1$, $H_0 = 0$, $V_r = 0.65$, $\alpha = 2 / \pi$,
$\Delta a_0 = 1.34$. 
Notably, this simulation cannot be continued beyond a finite time: the
vesicle evolves towards a configuration in which the membrane is about to
break up, which lies beyond the topological capability of the present
fixed-connectivity mesh, and the computation therefore stops.
Figure~\ref{fig:shear_ade} shows a sequence of vesicle shapes under shear
flow, ending with the last frame before this occurs. As the
shear flow is applied, the vesicle gradually tilts and elongates along the
flow direction, while the engulfed inner cavity is progressively stretched
into a slender pouch connected to the outer membrane through a thin neck. 
As shown in the final frame, this neck becomes increasingly thin and
exhibits a clear tendency toward rupture. Inspection of the vesicle
configuration at this instant confirms that the termination reflects a
genuine physical process: the imminent pinch-off of the neck, rather
than a numerical instability.

This observation allows us to anticipate a possible route for generating a
special class of vesicles. By tailoring the membrane properties through a
dedicated procedure, one may first drive the vesicle into the stomatocyte
configuration shown in Fig.~\ref{fig:steady_nonlocal}(b), and then impose a shear
flow to detach the engulfed inner portion from the outer membrane. This
would produce a distinctive double-membrane vesicle in which a large
vesicle encapsulates a smaller one, with the fluid enclosed in the inner
vesicle differing in species from the surrounding intermediate fluid,
thereby offering potential for specialized applications.

Furthermore, this points to a promising direction for the future
development of our method: to more fully exploit the Lagrangian nature of
the particle-based description, or even to replace the triangulated mesh
with a purely particle-based representation of the membrane. Such an
approach would enable a more faithful treatment of membrane rupture, and
could naturally capture processes such as membrane breakup, vesicle
endocytosis, budding, and the formation of new vesicles.

\section{\label{sec:conclusion}Conclusions}

In this work, we have proposed a novel numerical framework that couples a
multiphase smoothed particle hydrodynamics (SPH) method with a
triangulated-mesh discretization of vesicle membrane, and we have
applied it to study the equilibrium shapes and shear-flow dynamics of
vesicles governed by four different bending models.

The central feature of the proposed method is that the position of the
membrane mesh is determined by the interface between the inner and outer
fluid particles, identified through a color function, rather than by
direct velocity interpolation. This approach offers two distinct
advantages. First, the impermeability between the inner and outer fluids
is achieved naturally, so that non-physical constraints such as
particle reflection or bounce-back boundary conditions are not required to
prevent particles from crossing the membrane. Second, the enclosed volume
of the vesicle is preserved intrinsically through the weak compressibility
of the SPH formulation, eliminating the need for an artificial volume
constraint. In addition, the membrane model unifies four bending energy
formulations, namely the minimal, spontaneous-curvature (SC),
bilayer-couple (BC), and area-difference elasticity (ADE) models, within a
single framework, making the method versatile and comprehensive for diverse biophysical scenarios.

We first validated the method by reproducing the equilibrium shapes of
vesicles, including biconcave, prolate, dumbbell, asymmetric, and
stomatocyte morphologies, all of which agree with previously reported
results. We then simulated the tank-treading dynamics of a single vesicle
in shear flow using the minimal model, and obtained inclination angles and
revolution frequencies that are consistent with the theoretical
predictions of Kraus \textit{et al.} at large reduced volumes. The
deviations observed at small reduced volumes and low capillary number were
shown to originate from the non-ellipsoidal, centrally indented shapes
adopted by the vesicle, which fall outside the assumptions of the theory.

By comparing the four bending models, we found that the spontaneous
curvature has only a negligible effect on the steady orientation of the
vesicle, which is governed primarily by the geometric anisotropy set by
the reduced volume. In contrast, the revolution frequency is strongly
model-dependent at low reduced volumes: the SC model develops deeply
necked, dumbbell-shaped morphologies in which a substantial fraction of
the membrane circulates only within a single lobe, reducing the effective
radius of circulation and increasing the average angular velocity. The
non-local BC model instead favors laterally extended, grooved, and
asymmetric shapes, while the ADE model produces stomatocyte-like
configurations. For the stomatocyte vesicle under shear flow, the
simulation predicts a pronounced thinning of the neck and a clear tendency
toward rupture, suggesting a possible route for generating
vesicle-in-vesicle structures in which a large vesicle encapsulates a
smaller one filled with a different fluid.

These results demonstrate that the proposed SPH--mesh coupling provides an
accurate, robust, and comprehensive tool for investigating the dynamics of
vesicles with complex shapes in fluid flows. Several directions remain open
for future work. The current scheme cannot yet handle topological changes
such as membrane rupture, budding, and fission. A promising extension is
to more fully exploit the Lagrangian nature of the particle-based
description, or even to replace the triangulated mesh with a purely
particle-based membrane representation, which would enable a more faithful
treatment of membrane breakup, vesicle endocytosis, budding, and the
formation of new vesicles. Extending the method to more complex flow
configurations, to suspensions of multiple interacting vesicles, and to
higher Reynolds numbers would further broaden its range of applications.

\section*{Author Contributions}
Kuiliang Wang: methodology, software, investigation, writing -- original
draft. Xinwei Cai: software, validation. Ting Ye: conceptualisation,
review and editing. Xuejin Li: conceptualisation, review and editing. Xin Bian: conceptualisation, supervision, funding acquisition, writing, review and editing.

\section*{Acknowledgments}
This work is supported by the
National Key R\&D Program of China, Grant No. 2022YFA1203200 and National Natural Science Foundation of China, Grant No. 12172330.

\section*{Conflict of Interest}
The authors declare no potential conflict of interest.

\section*{Data Availability Statement}
The data that support the findings of this study are available from the
corresponding author upon reasonable request.


\appendix

\section{Particle distribution}
\label{app:particles}

\begin{figure}[htbp]
    \centering
    \begin{subfigure}[b]{0.48\textwidth}
        \centering
        \includegraphics[width=\textwidth]{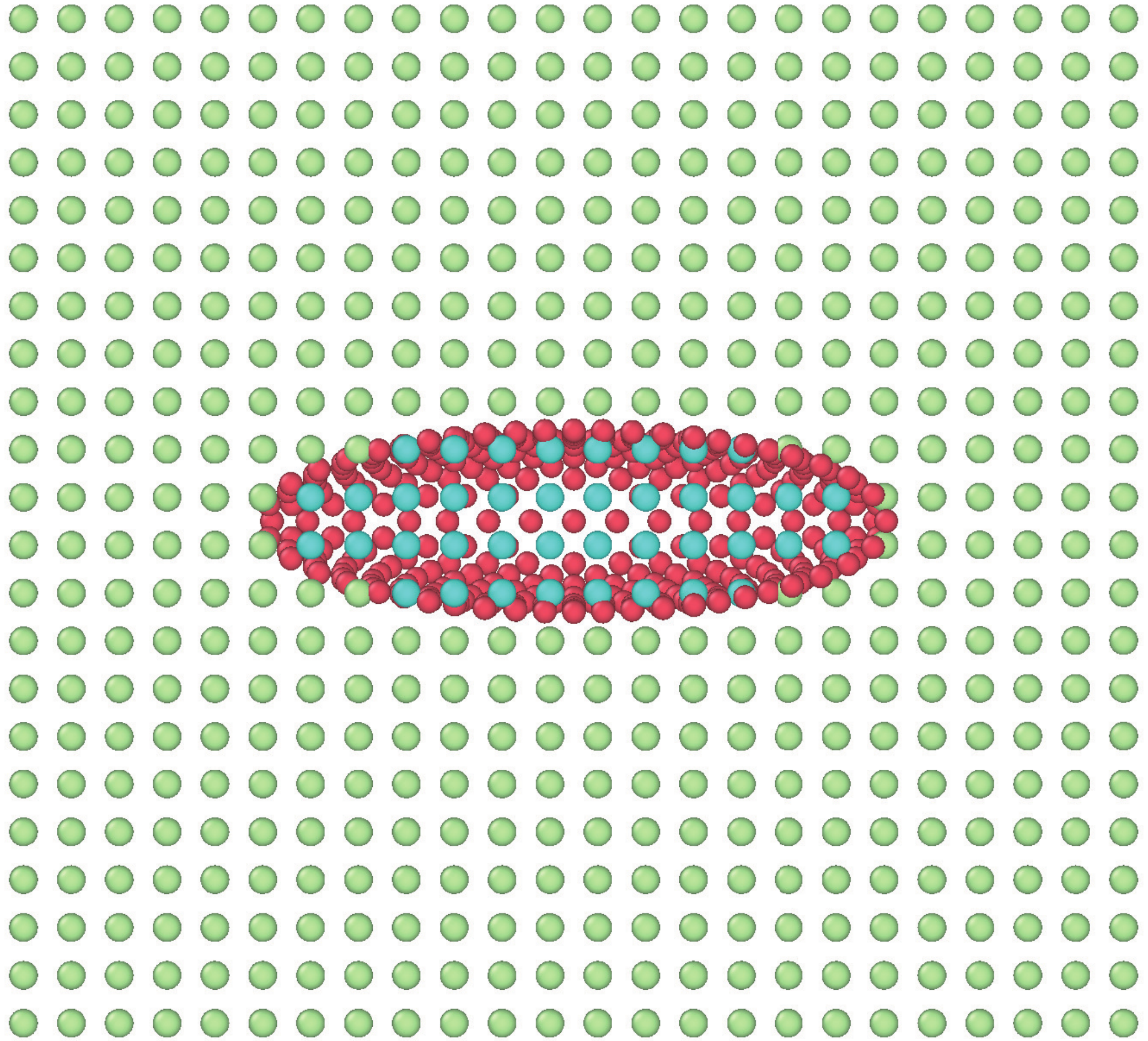}
        \caption{}
    \end{subfigure}
    \hfill
    \begin{subfigure}[b]{0.48\textwidth}
        \centering
        \includegraphics[width=\textwidth]{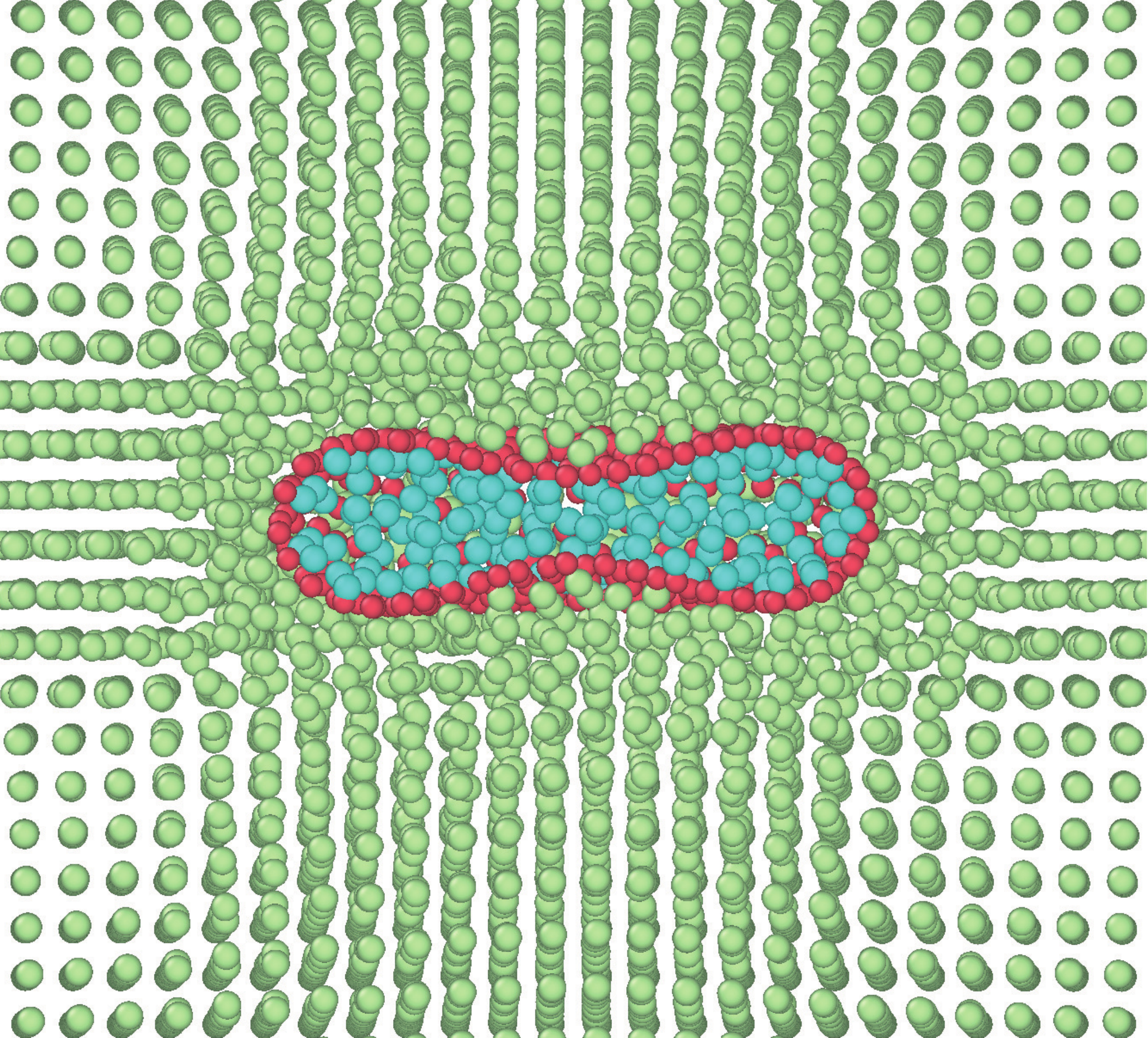}
        \caption{}
    \end{subfigure}
    \caption{Particle distribution during the energy-minimization process
    of a vesicle with $V_r = 0.65$ using the minimal model. (a) The initial
    oblate vesicle in a quiescent fluid. (b) The equilibrium biconcave
    shape. The view is a slice at $z = 0$.}
    \label{fig:minimization_particle}
\end{figure}

Figure~\ref{fig:minimization_particle} shows the particle distribution on the
mid-plane $z=0$ during the relaxation of a vesicle from an initial oblate
towards its energy-minimizing biconcave shape. The fluid particles inside and
outside the membrane redistribute smoothly as the shape evolves.

\clearpage

\bibliographystyle{unsrtnat}
\bibliography{wileyNJD-Chicago}

\end{document}